\documentclass[11pt]{article}
\usepackage{graphicx}
\usepackage{amssymb}
\usepackage{amsmath}
\usepackage{epsfig}
\usepackage{hyperref}
\usepackage{multirow}

\usepackage{epsf}
\usepackage{dcolumn}

\newcommand{\be}{\begin{equation}}
\newcommand{\ee}{\end{equation}}

\def\etal{{\it et al.},}

\newcommand\la{\lower0.6ex\vbox{\hbox{\ensuremath{\buildrel{\textstyle<}\over{\sim}\ }}}}
\newcommand\ga{\lower0.6ex\vbox{\hbox{\ensuremath{\buildrel{\textstyle>}\over{\sim}\ }}}}
\newcommand{\roughly}[1]%
    {{\mathrel{\raise.3ex\hbox{$#1$\kern-.75em\lower1ex\hbox{$\sim$}}}}}
\newcommand{\BM}[1]{{\mbox{\boldmath{$#1$}}}}

\def\lsim{\mathrel{\raise.3ex\hbox{$<$\kern-.75em\lower1ex\hbox{$\sim$}}}}
\def\gsim{\mathrel{\raise.3ex\hbox{$>$\kern-.75em\lower1ex\hbox{$\sim$}}}}

\begin{document}

\title{Generic dark matter signature for gamma-ray telescopes}
\author{V.~Barger$^{1}$, Y.~Gao$^{1}$,  W.-Y. Keung$^2$, D.~Marfatia$^{3}$\\[3ex]
\small\it $^1$Department of Physics, University of Wisconsin, Madison, WI 53706\\
\small\it $^2$Department of Physics,  University of Illinois, Chicago, IL 60607\\
\small\it  $^3$Department of Physics and Astronomy, University of Kansas, Lawrence, KS 66045
}

\date{}
\maketitle
\begin{abstract}
We describe a characteristic signature of dark matter (DM) annihilation or decay into gamma-rays. 
We show that if the total angular momentum of the initial DM particle(s) vanishes,
and helicity suppression operates to prevent annihilation/decay into light fermion pairs, then the amplitude for 
the dominant 3-body final state $f^+f^-\gamma$ has a unique form dictated by gauge invariance. 
This amplitude and the corresponding energy spectra hold for annihilation of DM Majorana fermions or self-conjugate scalars, and for decay of DM scalars, 
thus encompassing a variety of possibilities. Within this scenario, we analyze Fermi LAT, PAMELA and HESS data, and predict a hint in future Fermi gamma-ray 
data that portends a striking signal at atmospheric Cherenkov telescopes (ACTs).
  
\end{abstract}
\maketitle
\newpage

Hopes are high that the long-standing mystery of what comprises the dark matter of our universe may be resolved in the coming years. Recent data
from the Large Area Telescope (LAT) of the Fermi Gamma-ray Space Telescope~\cite{fermi} and from the High Eneregy Stereoscopic System (HESS)~\cite{hess} 
do not corroborate the excess $e^++e^-$ flux in data from the  
Advanced Thin Ionization Calorimeter (ATIC)~\cite{atic} and the Polar Patrol Balloon and Balloon borne Electron Telescope with
 Scintillating fibers (PPB-BETS)~\cite{bets} between 200 and 800~GeV. Also, the LAT $\gamma-$ray observations are discrepant with 
Energetic Gamma Ray Experiment Telescope (EGRET) data~\cite{Strong:2005zx} in that they do not confirm the excess at mid-latitudes in the energy range 
$10-50$~GeV~\cite{glastpri}.
Nevertheless, an excess in positrons between 10 and 270~GeV in the Payload for Matter Antimatter Exploration and Light-nuclei Astrophysics (PAMELA) 
data~\cite{pamela,pamnew} may be consistent with the LAT observations. The Fermi and PAMELA data may find a common explanation in DM annihilation or decay.  
 If future data establish the DM origin, a major breakthrough will be accomplished.


When a spin-0 DM particle decays, or a pair of identical DM particles of spin-${1\over2}$ or spin-0 annihilate in the static limit,
the total angular momentum of the initial state configuration is $j=0$. Conservation of angular momentum suppresses light fermion pair 
final states like $e^+e^-$ and $q\bar q$ without a chirality flip. This suppression disappears if the final state contains 
an additional photon. Following Ref.~\cite{bergstrom}, we refer to the emission of this additional photon in the final state 
as {\it internal bremsstrahlung} (IB) to distinguish it 
from {\it external} bremsstrahlung that requires an interaction with an external electromagnetic field. IB is comprised of photons radiated from the 
external legs {\it i.e., final state radiation} (FSR), and from internal lines {\it i.e., virtual IB} (VIB){\footnote {By a standard abuse of language, FSR
is defined to be the leading logarithmic contribution of the photon splitting from the external lines and is sometimes erroneously 
considered gauge-independent. 
Since VIB is not gauge-independent by itself, non-leading logarithmic contributions from external-line-bremsstrahlung need to be incorporated to 
render IB gauge-independent.}}$^,${\footnote{The distinction between FSR and VIB is artificial, since in an appropriate gauge, the
photons can be thought to have been radiated only from the external legs. However, we use this jargon since
it is intuitive.}}.
In this letter, we show that the chirality conserving amplitude for transitions from a $j=0$ initial state to 
a $e^+e^-\gamma$ final state is given by a unique form governed by QED gauge invariance.  
It follows that for a wide class of DM candidates, a distinct
$\gamma-$ray signal is expected to accompany the $e^\pm$ flux.

\begin{figure*}[t]
\mbox{\ \ \ \ \ \includegraphics[width=1.75in]{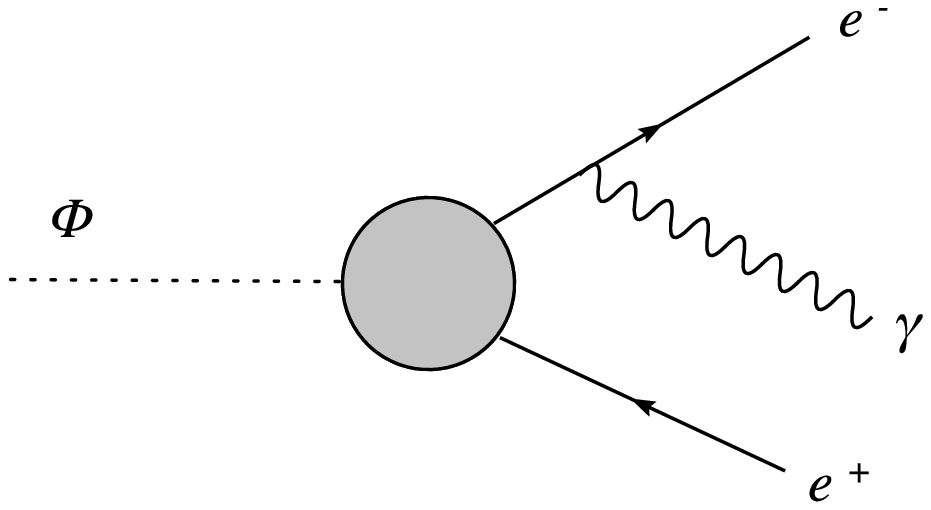}\ \ \ \ \ \ \ \ 
\includegraphics[width=1.75in]{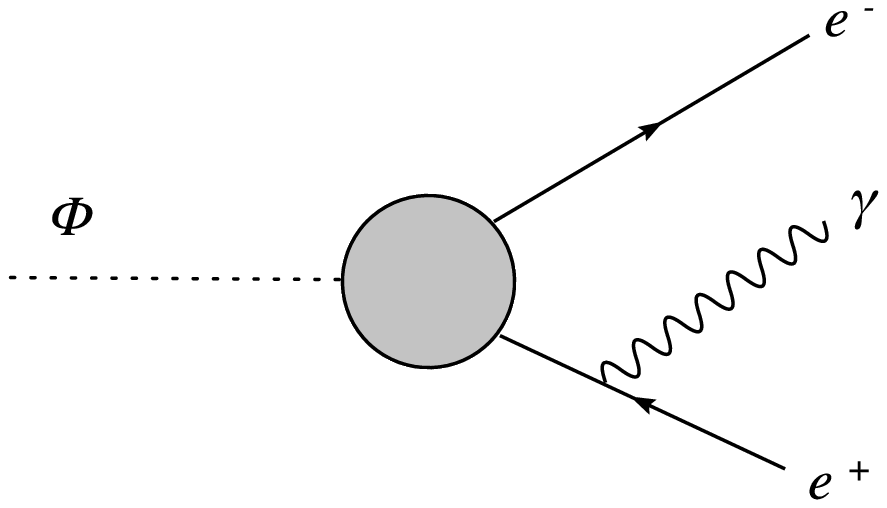}\ \ \ \ \ \ \ \
\includegraphics[width=1.75in]{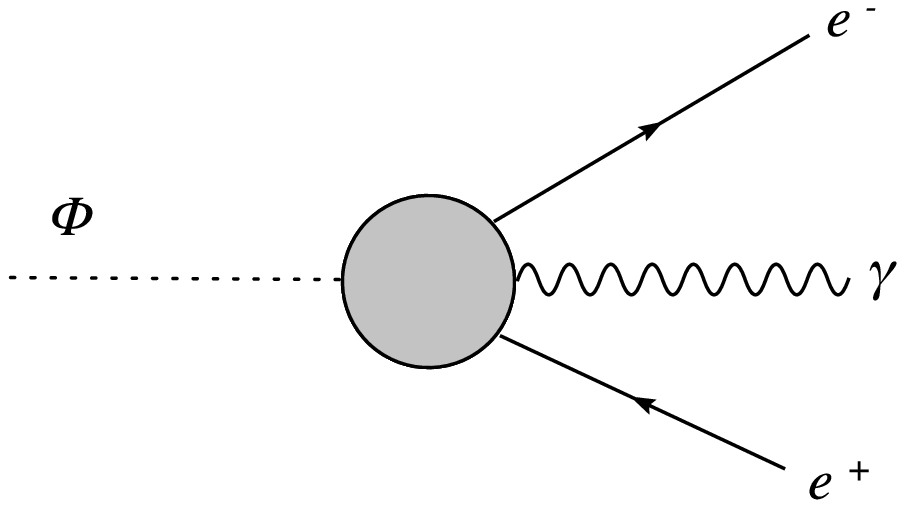}}                                  
\caption[]{Dominant diagrams for the chirality conserving process $\Phi \to e^+e^-\gamma$, where the initial scalar state 
$\Phi$ may be a decaying spin-0 DM particle or a pair of identical DM particles of spin-${1\over2}$ or spin-0 annihilating with zero total angular momentum.
\label{feynman}}
\end{figure*}

We are interested in the chirality preserving amplitude for the
final state $ e(p_1) +\bar e(p_2) +\gamma(k,\epsilon)$, from an
initial state of scalar structure: either a decaying DM
scalar boson, or a pair of identical DM particles 
with vanishing total angular momentum.
The QED gauge invariant amplitude must have
the form
\begin{equation}
 {\cal M} \sim
 \bar u_L(p_1)[ C(p_1,p_2) \not p_2\gamma_\mu \not k
          +C(p_2,p_1) \not k\gamma_\mu \not p_1 ] v_L(p_2) 
\epsilon^\mu + (L \to R)\,,
\label{unique}
\end{equation}
where the form function $C$ becomes a constant if the internal physics is from very short distances. 
Note that the  two terms do not interfere in the limit $m_e\to 0$.  
%
The operators corresponding to the amplitude with left chirality are 
\begin{eqnarray}
\Phi \bar \psi_{eL} \gamma_\mu (\partial_\nu \psi_{eL})  F^{\mu\nu} 
& \rightarrow  &
\bar u_L(p_1)(p_{2\mu}\not k-p_2\cdot k\gamma_\mu)  v_L(p_2)\epsilon^\mu \nonumber\\
   &=&  \hbox{$1\over2$} \bar u_L(p_1)\not p_2\gamma_\mu \not k  v_L(p_2)\epsilon^\mu\,,
\nonumber\\
\Phi (\partial_\nu\bar \psi_{eL}) \gamma_\mu  \psi_{eL}  F^{\mu\nu}
&\rightarrow&
   \hbox{$1\over2$} \bar u_L(p_1)\not k \gamma_\mu  \not p_1 v_L(p_2)\epsilon^\mu\,, \nonumber
\end{eqnarray}
where $\Phi$ is the initial scalar system with mass $M_\Phi = 2 m_{DM}$ for annihilation,
and $M_\Phi = m_{DM}$ for decay. The Feynman diagrams corresponding to the process are shown in Fig.~\ref{feynman}.

The rate for the annihilation process $\Phi \to e +\bar
e +\gamma$ is
$$ v_{\rm rel}{d \sigma\over dx_1 dz}
= {(y^2 e)^2\over 4 \pi^3 m_{DM}^2}  \    {[(1-x_1)^2+(1-x_2)^2] (1-z)
                              \over
                               (1-2x_1-r)^2  (1-2x_2-r)^2  } \ ,$$
where $y$ is the coupling of the intermediate particle of mass $m_E$ to electrons and the DM particle,{\footnote{As an example, 
consider a new sector that includes a left-right symmetric electroweak doublet of heavy leptons
$ L^T_{L,R}=(N^0, E^-)_{L,R}$, and a  gauge singlet scalar $\phi$, 
which is the DM particle.  Odd discrete parity is assigned to the new
particles, and even parity to SM particles.
The relevant interaction is 
${\cal L} \supset y  \overline{\ell_L} L_R  \phi$,
which only applies to the left-handed SM lepton doublet
$ \ell^T=(\nu , e^-)_L$.}}
 and $r=4m_E^2/M_\Phi^2$.
The scaling variables $x_i=2E_i/M_\Phi$ for the electron and positron, and \mbox{$z=2E_\gamma/M_\Phi$} are defined in the static center of mass frame
so that $x_1+x_2+z=2$. 

Hard photons arise primarily from VIB from a charged intermediate particle. 
The photon energy distribution is obtained
by integrating over $x_1\in (1-z,1)$~\cite{Flores:1989ru}:
\begin{equation}
 v_{\rm rel}{d \sigma\over dz}
={  (y^2 e)^2 \over 32\pi^3 m_{DM}^2} {1-z\over(1+r-z)^2} \bigg(   2 z{ z^2 + (1+r-z)^2 \over (1+r)(1+r-2z)} -{(1+r)(1+r-2z)\over 1+r-z}\ln {1+r\over 1+r-2z}\bigg)\,.\nonumber
\end{equation}

\begin{figure}[t]
\mbox{\includegraphics[width=6.5in]{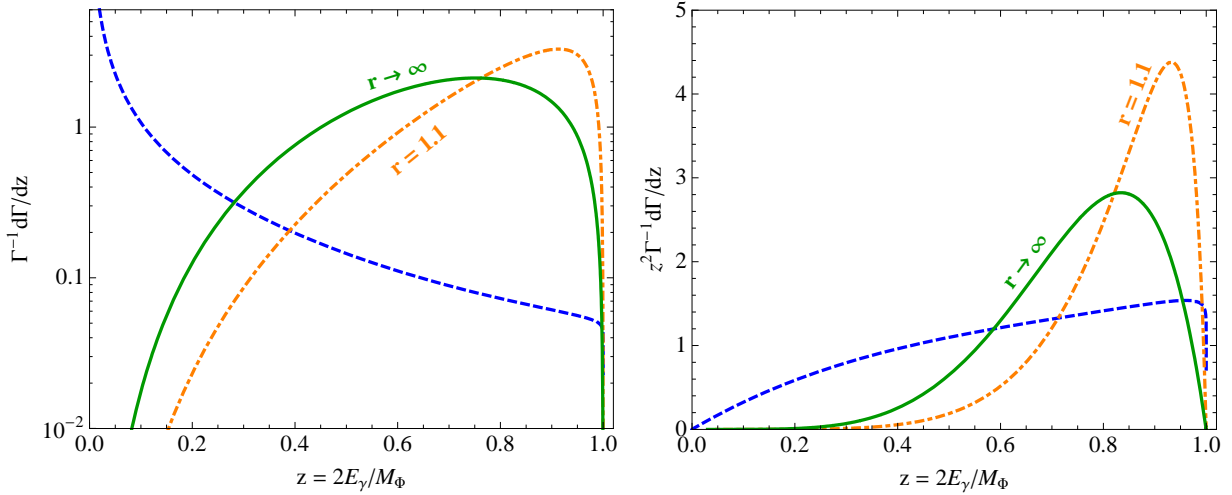}}                                      
\caption[]{Photon distributions including VIB compared with the distribution from FSR only (dashed curves). 
The $r \equiv 4m_E^2/M_\Phi^2\to \infty$ case is given by Eq.~(\ref{photondist}).
All distributions have unit area with the divergent FSR distribution in the left panel cut-off below 0.1~GeV for $M_\Phi=1$~TeV. Note the
 location of the peak in the FSR distribution in the two panels. The distributions in the
right panel are directly related to Fermi data which are presented as $E^2_\gamma \Phi_\gamma$. 
\label{dist}}
\end{figure}

If the exchanged particle is much heavier than the DM 
particle ($r \to \infty$), the relevant 
short distance physics scale $\Lambda\gg M_\Phi$ justifies the use
of a dimension-7 operator that is valid for both annihilation (with $M_\Phi = 2 m_{DM}$) and decay (with $M_\Phi = m_{DM}$):
$$  {e\over \Lambda_L^3} \Phi 
 \partial_\nu ( \bar \psi_{eL} \gamma_\mu\psi_{eL}) F^{\mu\nu}
+(L\to R) \ . $$
The differential decay distribution is
$$ {d\Gamma \over dx_1 dz}
={e^2 M_\Phi^7\over 512\pi^3 \Lambda_L^6}
(1-x_3)[(1-x_1)^2+(1-x_2)^2] + (L\to R)\,,$$
and the total width is
$$ \Gamma 
={e^2M_\Phi^7\over 15360\pi^3}
\left({1\over \Lambda_L^3}+{1\over \Lambda_R^3}\right)^2\,.
$$
The normalized distributions in $z$ of the prompt photon and in $x_2$ of the positron are
\begin{eqnarray}
{1\over \Gamma} {d\Gamma\over d z} &=&  20 (1-z) z^3\,, \label{photondist}\\
   {1\over \Gamma} {d\Gamma\over d x_2}
&=& 5(3-6x_2+\hbox{$7\over2$}x_2^2)x_2^2\,.
\end{eqnarray}
The decay distributions are the same as the normalized annihilation distributions in the short distance limit \mbox{$r\to \infty$}. 
(However, the experimental signatures for decay, and for annihilation with $r \to \infty$, are not identical because the injection flux
for decay depends on the DM halo distribution as $\rho$, while that for annihilation has a $\rho^2$ dependence.) 
The normalized photon distribution is shown in Fig.~\ref{dist}.
Note that $d\Gamma/dz$ increases from low $z$ to peak at
$z={3\over4}$ and then drops to zero.  On the other hand,
$d\Gamma/dx_2$ increases from
low $x_2$ of the positron and peaks at the endpoint $x_2=1$. 

To reproduce a lifetime of $10^{26}$~s for a DM particle with $M_\Phi \sim 2$~TeV as suggested by PAMELA, Fermi and HESS data, 
the typical short distance physics scale required is $\Lambda_L=\Lambda_R\sim 10^{11}$~GeV.

Our study is applicable to the annihilation of Majorana fermions (like neutralinos) and self-conjugate scalars, and
to decay of DM scalars~\cite{sterile}. Scalar DM annihilation through Higgs exchange does not fall within its purview because helicity suppression is not 
operative; we emphasize that the corresponding distributions will be markedly different. 
Incidentally, in models with spin-1 DM like minimal Universal Extra Dimensions (mUED)~\cite{mued} and Little Higgs with 
T-parity (LHT)~\cite{lht}, the photon energy distribution (accounting for VIB) is very flat and drops off precipitously at $m_{DM}$~\cite{bergs}. 

We simulate the spectra using GALPROP~\cite{galprop} as described in the appendix of Ref.~\cite{pap}. The default set-up in GALPROP produces
$\bar{p}/p$ spectrum that agrees with PAMELA data~\cite{antiprotons} above 10~GeV for which effects of solar modulation are insignificant. 
We consider the Einasto~\cite{Navarro:2008kc} and the isothermal~\cite{isoT} DM halo profiles to be representative of mildly cusped and cored 
profiles, respectively. 

\begin{figure*}[ht]
\mbox{\includegraphics[width=6.5in]{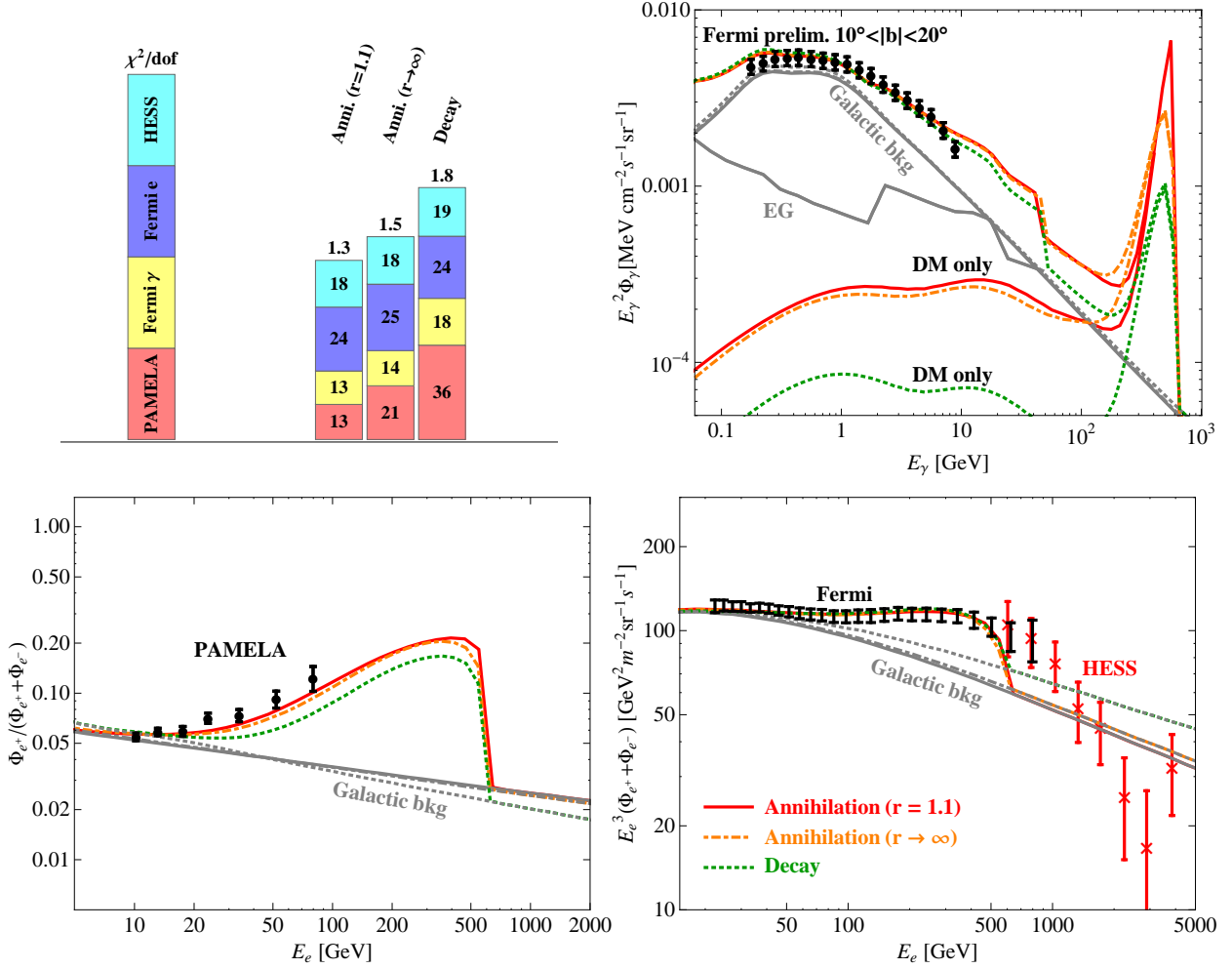}}                                      
\caption[]{DM annhilation and decay directly into $e^+e^-\gamma$ for $M_\Phi=1.2$~TeV.
 While ACTs will clearly see the bump structure in the $\gamma-$ray spectrum, Fermi may find
a flux enhancement at the upper limit of its sensitivity (300~GeV), but will not resolve the bump.
Best-fit $\chi^2$ values from a joint analysis of the PAMELA, Fermi $\gamma-$ray, Fermi $e^\pm$ and HESS datasets 
which have 7, 18, 26 and 8 points, respectively are provided in the top-left panel.
The number of free parameters is 8, and the number of degrees of freedom (dof) is 53, including two energy scale normalizations that account for
energy calibration uncertainties in the Fermi $e^{\pm}$ and HESS datasets. 
 The galactic background contribution which was fit for each case is displayed in the same line-type as for the signal. EG is the extragalactic 
$\gamma-$ray background that has been estimated up to about 50~GeV~\cite{reimer}.
 The HESS and Fermi error bars have been expanded to approximately include 
systematic uncertainties (apart from the energy scale uncertainties). See the appendix of Ref.~\cite{pap} for details of the statistical analysis. 
The DM halo follows the Einasto profile.
\label{fig:3}}
\end{figure*}

\begin{figure*}[t]
\mbox{\includegraphics[width=6.75in]{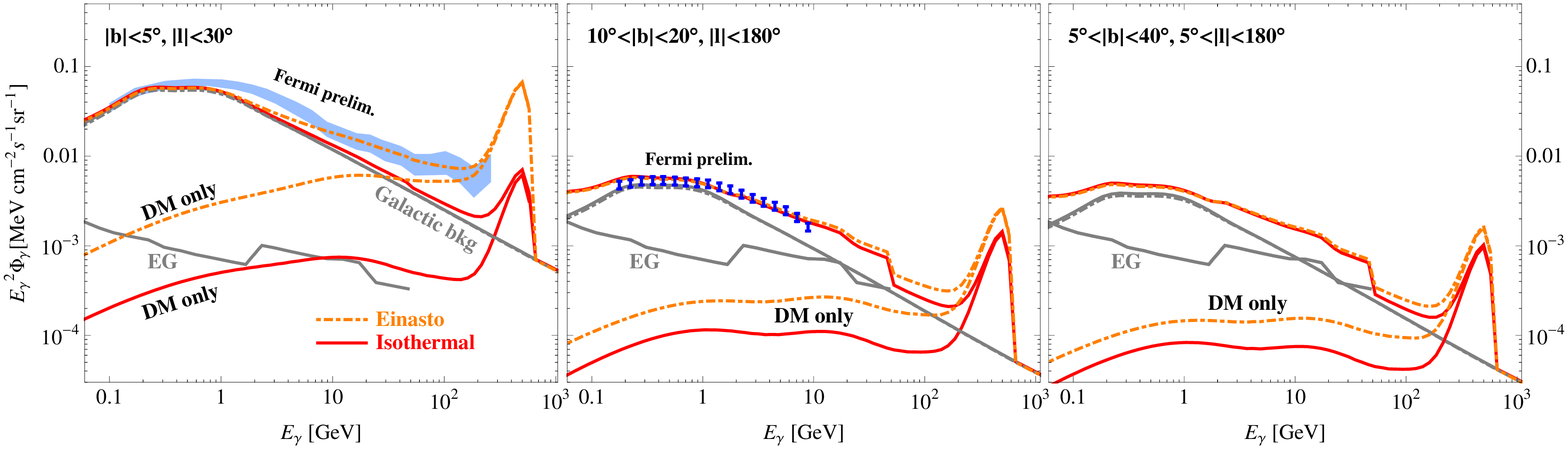}}
\caption[]{$\gamma-$ray signals for DM annihilation with $r\to \infty$ and $M_\Phi=1.2$~TeV for the Einasto and isothermal
profiles in three regions of the sky: inner galactic region, mid-latitudes, and
most of the sky outside the galactic center region. For both DM profiles, the signals correspond to spectra that fit current PAMELA, Fermi and HESS data;
in the left panel, very preliminary Fermi data~\cite{drell} are shown for comparison.
\label{fig:4}}
\end{figure*}

In Fig.~\ref{fig:3} we show that DM annihilation or decay dominantly into $e^+ e^-\gamma$ for $M_\Phi=1.2$~TeV 
easily explains the Fermi LAT, PAMELA and HESS data; larger values of $M_\Phi$ do not satisfactorily 
reproduce the steeply rising PAMELA positron fraction, and smaller values do not fit the HESS data.
 For DM annihilation, the $e^\pm$ flux
needs to be boosted by about 100 assuming a typical thermally averaged annihilation cross section 
$\langle \sigma  v\rangle = 3\times 10^{-26}$ cm$^3$s$^{-1}$ required to produce the measured relic abundance. This boost factor is an order of magnitude
more than expected from N-body simulations~\cite{Lavalle:1900wn}. For the decay case the lifetime is about $7\times 10^{26}$~s. 
Note the significant rise in the $\gamma-$ray flux at high energies that resembles line 
emission. The large amplitude of the signal distinguishes our class of scenarios from models which produce line emission at loop order. 
Examples with loop-dominated processes include models in which annihilation occurs through $s$-channel Higgs exchange as in several scalar DM models, 
mUED~\cite{mued1} and LHT~\cite{lht1}.  Another characteristic is the single large bump; lines from higher order diagrams may appear on the large bump with
much smaller amplitudes, and may not be resolvable. This is in contrast to models that produce multiple lines of roughly equal 
amplitude~\cite{forest}, as in theories with two universal extra dimensions compactified on a chiral square~\cite{dobrescu}. 
Future Fermi data may be able to see a flux enhancement with its 10\% energy resolution, 
but will not resolve the bump because it has limited sensitivity to photons with energy above 300~GeV. However, 
ACTs like 
the CANGAROO III system (Collaboration of Australia and Nippon for a Gamma Ray Observatory in the Outback), Major Atmospheric Gamma-ray Imaging
Cherenkov Telescope (MAGIC), the Very Energetic Radiation Imaging Telescope Imaging System (VERITAS), the Cherenkov Telescope Array (CTA),
 and the Advanced Gamma Ray Imaging System (AGIS)
which can detect photons in the 50~GeV to 100~TeV
range, will confirm the bump structure.
In Fig.~\ref{fig:4}, we forecast signals for these telescopes in different regions of the sky for the two DM halo profiles. We do not 
consider regions very close to the galactic center because the halo profile in the central region is affected by several astrophysical processes, 
that although model-dependent, tend to lower the DM density~\cite{bertone}. A significant bump
is seen in all regions of the sky for cored and cusped profiles.

It is known that annihilation of DM Majorana fermions into $e^\pm$ is enhanced with a concomitant sharp rise in the $\gamma-$ ray flux close to $m_{DM}$ 
when electromganetic radiative corrections that
relax helicity suppression are taken into account~\cite{bergstrom}. We have shown that the result holds for all scenarios in which 
the total angular momentum of the initial dark matter particle(s) vanishes,
and helicity suppression operates to prevent annihilation/decay into light fermion pairs, thus applying to scalar DM decay, and
to static annihilation of identical fermion or scalar DM particles.
We also found that the chirality preserving amplitude from the initial scalar state to $f^+f^-\gamma$ has the unique form in Eq.~(\ref{unique}) 
dictated by gauge invariance. 
The experimental signature for a wide class of DM models and candidates is unmistakable: a large bump in the high energy $\gamma-$ray flux that 
will be detectable by ACTs in any part of the sky. If extant PAMELA, Fermi $e^\pm$ and HESS data are to find an explanation in DM annihilation or decay,
Fermi may see the rising part of the bump. However, if $M_{\Phi}$ is slightly larger than 1.2~TeV, 
it is entirely possible that Fermi will not see a flux enhancement because of its 
insensitivity to photons above 300~GeV.

\vskip 0.1in
{\it Acknowledgments.}
This research was supported
by DOE Grant Nos.~DE-FG02-04ER41308,  DE-FG02-95ER40896 and DE-FG02-84ER40173,
by NSF Grant No.~PHY-0544278, and by the WARF.


\end{document}